\documentclass[12pt]{amsart} 
\usepackage[mathscr]{eucal}
\usepackage{amsmath,amsfonts}

\parskip=\smallskipamount

\newtheorem{theorem}{Theorem}

\newtheorem{example}[theorem]{Example}
\newtheorem{algorithm}[theorem]{Algorithm}

\makeatletter
\@addtoreset{equation}{section}
\makeatother

\newcommand{\cA}{{\mathcal A}}

\newcommand{\cD}{{\mathcal D}}

\newcommand{\cF}{{\mathcal F}}

\newcommand{\cH}{{\mathcal H}}
\newcommand{\cK}{{\mathcal K}}
\newcommand{\cL}{{\mathcal L}}
\newcommand{\cM}{{\mathcal M}}

\newcommand{\ZZ}{{\mathbb Z}}

\newdimen\expt
\def\boxit#1{\setbox0\hbox{$\displaystyle{#1}$}
      \hbox{\lower.4\expt
 \hbox{\lower3\expt\hbox{\lower\dp0
      \hbox{\vbox{\hrule height.4\expt
 \hbox{\vrule width.4\expt\hskip3\expt
      \vbox{\vskip3\expt\box0\vskip2\expt}%
 \hskip3\expt\vrule width.4\expt}\hrule height.4\expt}}}}}}

\begin{document}
\pagestyle{plain}

\bigskip

\title 
{Parametrizing quantum states and channels}
\author{T. Constantinescu} \author{V. Ramakrishna}

\address{Department of Mathematics \\
  University of Texas at Dallas \\
  Box 830688, Richardson, TX 75083-0688, U. S. A.}
\email{\tt tiberiu,vish@utdallas.edu}

\begin{abstract}
This work describes one parametrization
of quantum states and channels and
several of its possible applications. This parametrization works in any
dimension and there is an explicit algorithm which produces it.
Included in the list of applications are a simple characterization
of pure states, an explicit formula for one additive 
entropic quantity which does not require knowledge of
 eigenvalues, and an algorithm which finds one
Kraus operator representation for a quantum operation without recourse to
eigenvalue and eigenvector calculations.
\end{abstract}

\maketitle

\section{Introduction}

The interest in quantum information processing has brought added attention
to questions of parametrizations of positive matrices.
Indeed, at least two foundational ingredients in the theory of quantum
information, viz., quantum states and quantum channels 
involve positive matrices. See, for instance, \cite{nie,preskill}. 
Therefore it appears to be of interest to obtain descriptions 
(parametrizations) of the set of positive matrices.

The purpose of this note is to provide such an explicit parametrization
by adapting similar results on positive definite kernels in \cite{Co}.
This parametrization is rather intricate (and nonlinear), but in this
paper we show some benefits of its use: 
the parameters of pure states are
easily described;  the parameters of a tensor product can be deduced from
those of its factors; 
at least one entropic quantity (to be described in Section~3.3) is easy to
compute;  the Peres- Horodecki criterion for separability
can be described explicitly
described as inequalities for these parameters in low dimensions;
purifications of qubits can be explicitly parametrized, and finally one
Kraus operator
representation for quantum operations can be computed without requiring
any knowledge of the eigenvalues/eigenvectors of the associated Choi matrix.

The key features of this parametrization which are worth 
emphasizing are that it works in any dimension (indeed, it applies to matrices
whose entries are operators themselves); there are explicit formulae
for the parameters in terms of the entries of the matrix;
while these are rather intricate, there is one 
computationally attractive 
algorithm which produces this parametrization ; several quantities
of interest can be computed via these parameters; the algorithm exploits
the fact that the inherent structure in the matrix is inherited by its
Schur complements and this results in the algorithm yielding (at least) one
Cholesky decomposition of the matrix. Indeed, this last feature is 
precisely the reason why a Kraus operator representation of a quantum 
channel is produced by these parameters, without any need for eigenvalues
or eigenvectors.  

This paper is organized as follows.
After preliminaries concerning definitions, the main result
on the parametrization 
of positive semidefinite matrices is stated and illustrated via some
examples in the next section. 
We deal with states in Section~3 and 
with quantum channels in Section~4. An Appendix contains the sketch of
a proof of Theorem~1 on the parametrization of positive matrices 
and describes how to view a positive matrix as a matrix with a so-called
displacement structure and then describes an algorithm for the computation
of the parameters.

\section{Preliminaries}
In this section we introduce terminology and state the main result
concerning the 
parametrization of positive (semidefinite) matrices.

\noindent
{\it 2.1. Quantum states and channels.}\,
The state of a $d$-dimensional quantum system is described by 
a $d\times d$ positive density matrix of trace 1, that is, a
positive element of trace 1 in the algebra $\cM _d$ of complex $d\times d$ 
matrices. States described by rank one density matrices are called pure 
states.

A quantum channel is a completely positive map
$\Phi :\cA \rightarrow \cL({\cH})$
from a $C^*$-algebra $\cA $ into the set $\cL(\cH)$ of all bounded 
linear operators on the Hilbert space $\cH$ (in the situations most
frequently met in quantum information processing, $\cA = \cM _d$
and $\cL(\cH) = \cM _{d^{'}}$, while $\Phi$ is also required to
be trace preserving). 
By the Stinespring theorem, \cite{Pa}, Theorem~4.1, such a map 
is the compression of a $*$-homomorphism. For $\cA =\cM _d$, 
there is a somewhat more explicit
representation, given in \cite{Ch} (see also \cite{Jam}).
Thus, $\Phi :\cM _d\rightarrow
\cL(\cH)$ is completely positive if and only if
the matrix
\begin{equation}\label{sphi}
 S=S_{\Phi }=\left[\Phi (E_{k,j})\right]_{k,j=1}^d
 \end{equation}
 is positive, where $E_{k,j}$, $k,j=1,\ldots ,d,$
 are the standard matrix units of $\cM_d$. Each 
 $E_{k,j}$ is a $d\times d$ matrix
consisting 
of $1$ in the $(k,j)th$ entry and zeros elsewhere. 
 We notice that if $X=\left[X_{k,j}\right]_{k,j=1}^d$,
 then $Y=\left[Y_{k,j}\right]_{k,j=1}^d=\Phi (X)$ is given by the relations
 \begin{equation}\label{filter}
 Y_{k,j}=\sum _{l,m}\Phi (E_{l,m})_{k,j}X_{l,m}.
 \end{equation} 
This shows that there is a one-to-one
correspondence between the set of completely positive maps on 
$\cM _d$ with values in $\cL (\cH )$ and the set of positive 
matrices in $\cM _d \otimes \cL (\cH )$.

For a linear map $\Phi :\cM _d\rightarrow \cM _d$ 
the adjoint $\hat \Phi $ is defined with respect
to the Hilbert space structure on $\cM _d$ given by the Hilbert-Schmidt
inner product $\langle A,B\rangle =Tr(A^*B)$, where $A^*$ denotes the 
adjoint of $A$. It is easily seen that $\Phi $ is trace preserving 
if and only if $\hat \Phi $ is unital ($\hat \Phi (I)=I$).

\noindent
{\it 2.2. A parametrization of positive matrices.}\,
We describe a parametrization of the positive matrices in 
 $\cM _d \otimes \cL (\cH )$, with $\cH$ allowed to be infinite-dimensional.
Note that if ${\mbox dim} (\cH ) = 1$, such
matrices are precisely $d\times d$ positive matrices with complex entries.
To that end, some elements of dilation theory (\cite{Pa}) are needed. 
Let $\cH _1$ and $\cH _2$ be two Hilbert spaces, not
necessarily finite dimensional, and let $\cL(\cH_1,\cH_2)$ denote the set
of all bounded linear maps operators from  $\cH_1$ into $\cH_2$.
The operator $T\in \cL(\cH_1,\cH_2)$ is called a contraction
if $\|T\|\leq 1$. The defect operator of $T$ is 
$D_T=(I-T^*T)^{1/2}$, where $T^*$ denotes the adjoint operator, 
(as well as the complex conjugate in case $T$ is just a complex number) 
and let $\cD_T$ denote the closure of the range
of $D_T$. To any contraction $T\in \cL(\cH_1,\cH_2)$
is associated the unitary operator $U(T):\cH_1\oplus \cD_{T^*}
 \rightarrow \cH_2\oplus \cD_{T}$ by the formula:
 \begin{equation}\label{rotele}
 U(T)=\left[
 \begin{array}{cc}
 T & D_{T^*} \\
 D_T & -T^* 
 \end{array}
 \right].
 \end{equation}

\vspace*{2mm}

\noindent Now let $\{\cF _k\}_{k=1}^d$ be a family of Hilbert spaces
and consider $\{\Gamma _{k,j}\mid k,j=1,\ldots ,d, k\leq j\}$ a family
of contractions such that $(i)$ $\Gamma _{k,k}=0$ for $k=1,\ldots ,d$
and $(ii)$ for $k<j$, 
\begin{equation}\label{boundary}
\Gamma _{k,j}\in \cL (\cD _{\Gamma _{k+1,j}}, 
\cD _{\Gamma ^*_{k,j-1}}). 
\end{equation}
Unitary operators $U_{k,j}$ are associated
to this family by the recursions:
$U_{k,k}=I_{\cF _k}$, the identity operator on $\cF _k$, $k=1,\ldots ,d$,
while  for $j>k$,
 $$U_{k,j}=U_{j-k}(\Gamma _{k,k+1})U_{j-k}(\Gamma _{k,k+2})\ldots
 U_{j-k}(\Gamma _{k,j})(U_{k+1,j}\oplus I_{\cD _{\Gamma ^*_{k,j}}}),
$$
where $U_{j-k}(\Gamma _{k,k+l})$ denotes the unitary 
operator defined from the space 
$$
\left(\oplus _{m=1}^{l-1}\cD _{\Gamma _{k+1,k+m}}
 \right)\oplus (\cD _{\Gamma _{k+1,k+l}}
 \oplus \cD _{\Gamma ^*_{k,k+l}})\oplus
 \left(\oplus _{m=l+1}^j\cD _{\Gamma _{k,k+m}}\right)
$$
onto the space 
$$
\left(\oplus _{m=1}^{l-1}\cD _{\Gamma _{k+1,k+m}}
 \right)\oplus (\cD _{\Gamma ^*_{k,k+l-1}}
 \oplus \cD _{\Gamma _{k,k+l}})\oplus
 \left(\oplus _{m=l+1}^j\cD _{\Gamma _{k,k+m}}\right)
$$
 by the formula
$$
U_{j-k}(\Gamma _{k,k+l})=I\oplus U(\Gamma _{k,k+l})\oplus I.
$$

\vspace*{2mm}

\noindent The foregoing considerations from dilation theory will now
be applied to matrices. 
Let $S=\left[S_{k,j}\right]_{k,j=1}^d$ be a matrix such that 
$S_{k,j}\in \cL (\cH )$ and let $S_{k,k}=L_{k,k}^*L_{k,k}$ be a
factorization of $S_{k,k}$, $k=1,\ldots ,d$. 
Denote the closure of the range of $ L_{k,k}$ by $\cF _k$, 
$k=1, \ldots ,d$. For a family of contractions $\Gamma_{kj}$ as in
the previous paragraph, denote by $R_{k,j}$ the row contraction 
$$
R_{k,j}=\left[
 \begin{array}{cccc}
 \Gamma _{k,k+1}, & D_{\Gamma ^*_{k,k+1}}\Gamma _{k,k+2},  
 & \ldots ,& D_{\Gamma ^*_{k,k+1}}\ldots D_{\Gamma ^*_{k,j-1}}\Gamma _{k,j}
 \end{array}\right]
$$
and by $C_{k,j}$  the column contraction 
$$
C_{k,j}=\left[
 \begin{array}{cccc}
 \Gamma _{j-1,j}, & \Gamma _{j-2,j}D_{\Gamma _{j-1,j}},  
 & \ldots ,& \Gamma _{k,j}D_{\Gamma _{k+1,j}}\ldots D_{\Gamma _{j-1,j}}
 \end{array}\right]^t,
$$
 where $"t"$ stands for matrix transpose.
We now obtain the following characterization and structure
of a positive matrix.

\begin{theorem}\label{param}
The matrix $S=\left[S_{k,j}\right]_{k,j=1}^d$ 
as above, satisfying $S_{jk}^{*} = S_{kj}$, is positive if and only if
i) $S_{kk} \geq 0, k=1, \ldots, d$ and ii) there exists a family 
$\{\Gamma _{k,j}\mid k,j=1,\ldots ,d, k\leq j\}$ of contractions
such that  
$\Gamma _{k,k}=0$ for $k=1,\ldots ,d$ and Equation (\ref{boundary}) is
valid, and  
\begin{equation}\label{relation}
 S_{k,j}=L^*_{k,k}(R_{k,j-1}U_{k+1,j-1}C_{k+1,j}+
 D_{\Gamma ^*_{k,k+1}}\ldots 
 D_{\Gamma ^*_{k,j-1}}\Gamma _{k,j}
 D_{\Gamma _{k+1,j}}\ldots
 D_{\Gamma _{j-1,j}})L_{j,j}.
 \end{equation}
\end{theorem}

For a proof see Appendix ~1.
{\it The $\Gamma_{kj}, j\neq k$ form
the parametrization of positive matrices referred
to in Section 1}. Note the $\Gamma_{kk}$ are just some fake parameters,
included in the statement of the theorem to avoid an artificial separation
of the $j = k+1$ case from that for other values of $j$ 
Note that when the dimension of $\cH = 1$, these 
contractions, $\Gamma_{kj}$, are complex numbers in the closed unit disc) .

Though this result looks rather intricate 
it provides a true parametrization 
of the set of positive matrices, has a useful physical interpretation, 
and does not depend on the dimension of $\cH$. Furthermore, by adopting
certain natural conventions,
this can be even turned into a one-one parametrization
of quantum states. One illustration of this is provided in Section 3, for the
$d=2$ case.

Let us next look at some concrete examples of Theorem 1. 
For $d=2$ the result is well-known (see, e.g. \cite{Pa}). For 
$d=3$, the structure is more interesting, even in the scalar case
$\dim \cH =1$. Thus, let $S=\left[\begin{array}{ccc}
S_{11} & S_{12} & S_{13} \\
S^*_{12} & S_{22} & S_{23} \\
S^*_{13} & S^*_{23} & S_{33}
\end{array}\right]$
be a positive matrix. Theorem~\ref{param}
gives:
$$
S_{12}=L^*_{11}\Gamma _{12}L_{22},
$$
$$
S_{23}=L^*_{22}\Gamma _{23}L_{33},
$$
$$
S_{13}=L^*_{11}\left(\Gamma _{12}\Gamma _{23}+
D_{\Gamma ^*_{12}}\Gamma _{13}
D_{\Gamma _{23}}\right)L_{33},
$$
and we can notice the connection between the structure of $S$
and spherical geometry in the formula for $S_{13}$. Thus, 
for $\Gamma _{12}=\cos \theta $, $\Gamma _{23}=\cos \theta _1$
and $\Gamma _{13}=\cos \phi$, the above formula for $S_{13}$
reduces to the cosine law in spherical geometry.

We also write the case $d=4$ explicitly, with $\cH $ of arbitrary 
dimension; in this case, 
if $S=\left[S_{k,j}\right]_{k,j=1}^4$, then
$$
S_{k,k+1}=L^*_{k,k}\Gamma _{k,k+1}L_{k+1,k+1}, \quad k=1,2,3,
$$
$$
S_{k,k+2}=L^*_{k,k}\left(\Gamma _{k,k+1}\Gamma _{k+1,k+2}
+D_{\Gamma ^*_{k,k+1}}\Gamma _{k,k+2}
D_{\Gamma _{k+1,k+2}}\right)L_{k+2,k+2},
\quad k=1,2,
$$
and 
$$
\begin{array}{rcl}
S_{14}&=&L^*_{11}\left(
\Gamma _{12}\Gamma _{23}\Gamma _{34}+
D_{\Gamma ^*_{12}}\Gamma _{13}D_{\Gamma _{23}}\Gamma _{34}
+\Gamma _{12}D_{\Gamma ^*_{23}}\Gamma _{24}D_{\Gamma _{34}}\right. \\
 & & \\
 & & \quad \left.-D_{\Gamma ^*_{12}}\Gamma _{13}\Gamma ^*_{23}
\Gamma _{24}D_{\Gamma _{34}}
+D_{\Gamma ^*_{12}}D_{\Gamma ^*_{13}}
\Gamma _{14}D_{\Gamma _{24}}D_{\Gamma _{34}}\right)L_{44}.
\end{array}
$$

\noindent
{\it Lattice Structures and Time-Dependent String Models}
It is worthwhile to digress briefly to point out that in the Appendix,
positive matrices are shown to be a special case of matrices with 
displacement structure, and that this results in an algorithm for the
calculation of the parameters $\Gamma_{j,k}$. Furthermore, due to the
special choice of a lower triangular $F(-t)$ (in the
definition of a displacement structure) made in the appendix,
this algorithm has
the so-called lattice-structure.  
The same structure produces a Cholesky factorization of $S$,
$$
S=\left(\oplus _{k=1}^dL^*_{k,k}\right)
G^*_{1,d}G_{1,d}\left(\oplus _{k=1}^dL_{k,k}\right),
$$
where the upper-triangular operator $G_{1,d}$
is defined recursively by $G_{k,k}=I$, $k=1,\ldots ,d$, and 
for $1\leq k<j\leq d$, 
\begin{equation}\label{cholescu}
G_{k,j}=\left[\begin{array}{cc}
G_{k,j-1} & U_{k,j-1}C_{k,j} \\
0 & D_{\Gamma _{k,j}}\ldots D_{\Gamma _{j-1,j}}
\end{array}
\right].
\end{equation} 

These lattice structures can alternatively be described in the form 
of a time-varying, discrete transmission-line (string) 
For more details see \cite{Co},
where this connection is described in the context of displacement structures. 
In the invariant time case, this transmission
line is familiar in marine seismology, where the one-sided perfect reflector
is given by the interface air-water, see \cite{Cl}.
The implications of this classical model for quantum states remain to be
fully worked out. However, this model explains the intriguing
fact that the number
of summands in the formula for $S_{k, k+l}, l= 0, 1, \ldots $, in Theorem 1,
is precisely
the $l$th Catalan number. These observations will be explained elsewhere.

\section{Parametrization of states}
In this section we parametrize  (finite dimensional)
quantum states by using Theorem~\ref{param}
and we show several applications of this parametrization.
{\it Throughout this section, except in Subsection 3.2, 
we will write the $\Gamma_{k,l}$ and the $D_{\Gamma ^*_{k,l}}$ of the
previous section as $g_{k,l}$ and $d_{k, l}$ respectively, so as to
distinguish states from channels.}
The explicit formulae for the parameters, $g_{k,l}$, 
depend on the basis
of $\cM _d$ and it is often convenient to use
selfadjoint bases containing the identity (see, for instance, 
\cite{Za}, \cite{Ki}).
Such an orthogonal basis can be obtained from $\{E_{k,j}\}_{k,j=1}^d$
as follows:
$$f^d_{k,j}=E_{k,j}+E_{j,k},\quad k<j,$$
$$f^d_{k,j}=\frac{1}{i}\left(E_{j,k}-E_{k,j}\right),\quad k>j,$$
$$h_1^d=I_d,\quad h_k^d=h_k^{d-1}\oplus 0, \quad 1<k<d, \quad 
h_d^d=\sqrt{\frac{2}{d(d-1)}}\left(h_1^{d-1}\oplus (1-d)\right).$$
For $d=2$, we deduce
$$h_1^2=I_2, f_{12}^2=\sigma _1, f_{21}^2=\sigma _2, h_2^2=\sigma _3,$$
where $\sigma _1$, $\sigma _2$, $\sigma _3$
are the Pauli matrices. In the Pauli basis a hermitian matrix with 
trace $1$ will be 
representated as $\rho =\frac{1}{2}(I_2+\sum _{k=1}^3\beta _k\sigma _k)$
and its positivity is equivalent to the Bloch sphere condition
$1-\|\beta \|^2\geq 0$. By using Theorem~ \ref{param}, 
the positivity of $\rho $ is equivalent to a cylinder condition on 
the parameter $(\beta _3, g)$:
$$|\beta _3|\leq 1,$$
$$\beta _1-i\beta _2=(1-|\beta _3|^2)^{1/2}g, \quad |g|\leq 1,$$
and in order to ensure a one-to-one parametrization 
we choose $g=0$ if $|\beta _3|=1$.

For $d=3$ we obtain the Gell-Mann matrices:
$$h_1^3=I_3, \,\,\,h_2^3=\left[
\begin{array}{ccc}
 1 & 0 & 0 \\
 0 & -1 & 0 \\
 0 & 0 & 0 
\end{array}\right],\,\,\,
h_3^3=\frac{1}{\sqrt{3}}
\left[
\begin{array}{ccc}
 1 & 0 & 0 \\
 0 & 1 & 0 \\
 0 & 0 & -2 
\end{array}\right], \ldots .
$$
A hermitian matrix of trace $1$ will be represented as 
$$
\rho =\frac{1}{3}\left(I_3+\beta _2h_2^3+\beta _3h_3^3+
\sum _{k\ne j}\gamma _{k,j}f^3_{k,j}\right).
$$
By using Theorem~\ref{param}, the positivity of $\rho $ is 
equivalent to the conditions:
$$1+\beta _2+\frac{\beta _3}{\sqrt{3}}\geq 0,\quad 
1-\beta _2+\frac{\beta _3}{\sqrt{3}}\geq 0,\quad 
1-\frac{2\beta _3}{\sqrt{3}}\geq 0,$$
$$\gamma _{12}-i\gamma _{21}
=((1+\frac{\beta _3}{\sqrt{3}})^2-\beta _2^2)^{1/2}g_{12}, 
\quad |g_{12}|\leq 1,$$
$$\gamma _{23}-i\gamma _{32}
=
(1-\beta _2+\frac{\beta _3}{\sqrt{3}})^{1/2}
(1-\frac{2\beta _3}{\sqrt{3}})^{1/2}g_{23}, \quad |g_{23}|\leq 1,$$
$$\gamma _{13}-i\gamma _{31}
=(1-\beta _2+\frac{\beta _3}{\sqrt{3}})^{1/2}
(1-\frac{2\beta _3}{\sqrt{3}})^{1/2}
\left(g_{12}g_{23}+(1-|g_{12}|^2)^{1/2}(1-|g_{23}|^2)^{1/2}g_{13}\right), 
\quad |g_{13}|\leq 1.$$

The case $d=4$ was explicitly written in\cite{CR}, but with respect 
to the basis $\sigma _k\otimes \sigma _j$. In any dimension we can 
write 
$$\rho =\frac{1}{n}\left(I_n+\sum _{l=2}^d\beta _lh^d_l
+\sum _{k\ne j}\gamma _{k,j}f^d_{k,j}\right),$$
and the parameters $\beta _l$ appearing on the diagonal 
of $\rho $ satisfy a system of linear inequalities, while the 
parameters $g_{k,j}$ associated to the off-diagonal elements 
by Theorem~\ref{param}
are almost independent: they belong to the closed unit disk and satisfy 
the boundary conditions \eqref{boundary}.

\noindent
{\it 3.1. Pure states.}
It was already noticed in \cite{CR} that the purity of qubits can be 
explicitily checked via the parameters of Theorem~ \ref{param}.
Here we extend this result to any dimension.
Let $\rho $ be a $d$-dimensional state,
$$\rho =\frac{1}{n}\left(I_n+\sum _{l=2}^d\beta _lh^d_l
+\sum _{k\ne j}\gamma _{k,j}f^d_{k,j}\right)=
\frac{1}{n}\sum _{k,j}\rho _{k,j}E_{k,j}.$$
We notice that 
$$\rho _{k,k}=1+\sum _{l=2}^d\beta _l(h_l^d)_{k,k}, \quad k=1,\ldots ,d,$$
and 
$$
\rho _{k,j}=\gamma _{k,j}-i\gamma _{j,k}, \quad k<j.$$
Let $\{g_{k,j}\}$ be the parameters associated to the positive matrix
$\rho =\left[\rho _{k,j}\right]_{k,j=1}^d$ by Theorem~\ref{param}.

\begin{theorem}\label{purity}
A state $\rho $ is pure if and only if the parameters $g_{k,j}$
are zero except for those indices
for which $\rho _{k,k}\rho _{j,j}\ne 0$, in which case $|g_{k,j}|=1$.
\end{theorem}
\begin{proof}
Assume $\rho $ is pure. This implies that each matrix
$$\left[\begin{array}{cc}
\rho _{k,k} & \rho _{k,k+1} \\
\rho ^*_{k,k+1} & \rho _{k+1,k+1}
\end{array}
\right]
$$
has rank at most $1$. This happens in case either at least
one of $\rho _{k,k}$, $\rho _{k+1,k+1}$ is zero or if $|g_{k,k+1}|=1$.
It remains to see what happens if $\rho _{k,k}\ne 0$,
$\rho _{l,l}=0$ for $k<l<m$ and $\rho _{m,m}\ne 0$. In this situation
we must have $\rho _{p,l}=0$ for $k<p<l<m$, therefore
$g_{p,l}=0$ for $k<p<l<m$, and so, $\rho _{k,m}=
\rho _{k,k}^{1/2}g_{k,m}\rho _{m,m}^{1/2}$ 
by \eqref{relation}. Then, the fact that 
$$\left[\begin{array}{cc}
\rho _{k,k} & \rho _{k,m} \\
\rho ^*_{k,m} & \rho _{m,m}
\end{array}
\right]
$$
has rank at most $1$ implies $|g_{k,m}|=1$.

Conversely, we deduce from \eqref{relation} that 
$\rho =\frac{1}{n}v_{\rho }v^*_{\rho }$, where
the vector $v_{\rho }$ is described as follows:
let $i_1<\ldots <i_k$ be the indices of the nonzero diagonal elements
of $\rho $. Then, 
$$(v_{\rho })_{i_1}=\rho ^{1/2}_{i_1,i_1},$$
$$(v_{\rho })_{i_2}=\rho ^{1/2}_{i_2,i_2}g^*_{i_1,i_2},$$
$$\vdots $$
$$(v_{\rho })_{i_k}=\rho ^{1/2}_{i_k,i_k}g^*_{i_1,i_2}\ldots 
g^*_{i_{k-1},i_k},$$
while the other entries of $v_{\rho }$ are zero.
\end{proof}

Thus, the vector in $C^{d}$ representing the pure state can also be written
down in terms of the parameters.   
It was noticed in \cite{CR} that these parameters
can be used in producing a one-one parametrization of purifications of qubits.

\noindent
{\it 3.2. Tensor products.}
We next discuss producing parameters for tensor
products of positive matrices, in terms of the parameters of the factors
entering the product.

Let us begin with a very simple example. Take
$S_1=\left[\begin{array}{cc}
1 & A_{12} \cr
A^*_{12} & 1 
\end{array}
\right]$ 
and 
$S_2=\left[\begin{array}{cc}
1 & B_{12} \cr
B^*_{12} & 1 
\end{array}
\right]$.
Then 
$$S_1\otimes S_2=
\left[\begin{array}{cccc}
1& B_{12} & A_{12} & A_{12}B_{12} \cr
B^*_{12} & 1 & A_{12}B^*_{12} & A_{12} \cr
A^* _{12} & B_{12}A^* _{12} & 1 & B_{12} \cr
 B^*_{12}A^* _{12} & A^* _{12} & B^*_{12} & 1
\end{array}
\right]
$$ 
(note that we use the so-called right Kronecker 
product representation of the tensor product, see \cite{Be}).
Formula \eqref{relation} gives:

$$\Gamma _{12}=B_{12},\quad \Gamma _{23}=A_{12}B^*_{12},
\quad \Gamma _{34}=B_{12},$$
$$\Gamma _{13}=\frac{A_{12}(1-|B_{12}|^2)^{1/2}}
{(1-|A_{12}|^2|B_{12}|^2)^{1/2}},$$
$$\Gamma _{24}=\frac{A_{12}(1-|B_{12}|^2)^{1/2}}
{(1-|A_{12}|^2|B_{12}|^2)^{1/2}},$$
$$\Gamma _{14}=-A_{12}B_{12}.$$ 

The fact that 
$\Gamma _{14}$ has a simpler structure than expected indicates
that there is additional structure that can be explored fruitfully.  
It turns out that in this case it is more convenient
to search for a block lattice structure first.

Let $S_1=\left[S^1_{k,j}\right]_{k,j=1}^d$, 
$S_2=\left[S^2_{k,j}\right]_{k,j=1}^{d'}$
be two positive matrices. Also let 
$\{\Gamma ^1_{k,j}\}$,  $\{\Gamma ^2_{k,j}\}$ be the 
corresponding parameters associated by Theorem~\ref{param}
to $S_1$, respectively, $S_2$. For a matrix $T$ we use the notation
$T^{\oplus p}=\underbrace{T\oplus \ldots \oplus T}_{p\,\, times}$, to
denote its $p$-fold direct sum.

 \begin{theorem}\label{tensor}
The positive matrix $S_1\otimes S_2$ is a $d\times d$ block matrix
with structure
given by \eqref{relation} with:

$$L_{k,k}=G^2_{1,d'}(S^1_{,k,k})^{\oplus d'},\quad k=1,\ldots, d,$$
$$\Gamma _{k,j}=(\Gamma ^1_{k,j})^{\oplus d'},\quad 1\leq k<j\leq d,$$
where $G^2_{1,d'}$ is the upper triangular Cholesky factor of 
$S_2$.
 \end{theorem}

\begin{proof}
First, it is easily checked that
$$S_1\otimes S_2=(G^2_{1,d'})^{\oplus d})^*
\left[(S^1_{k,j})^{\oplus d'}\right]_{k,j=1}^d
(G^2_{1,d'})^{\oplus d}$$
and then, some algebra with formula \eqref{relation} 
gives that the parameters associated by Theorem ~\ref{param} to the 
positive matrix
$\left[(S^1_{k,j})^{\oplus d'}\right]_{k,j=1}^d$
are $\Gamma _{ij}=(\gamma _{ij}^1)^{\oplus M}$.
\end{proof}

It is quite convenient to use the Cholesky factor
of $S_2$ since it can be explicitly computed in 
terms of the parameters ${\Gamma ^2_{k,j}}$, as shown by  
\eqref{cholescu}.

\noindent
{\it 3.3. Entropy.}
In this section we consider an entropy-like number 
$E(\rho )$ that can be explicitly computed in terms of the parameters
$\{g_{k,j}\}$ of a state $\rho $. This quantity, $E(\rho ) =
\frac{1}{d}{\mbox Tr} \ (\log \rho)$, is mentioned in \cite{OP},
as one of a list
of candidates for a quantum notion of entropy. 
Further material about quantum entropies is summarized in \cite{OP}.

One simple way to motivate $E(\rho )$ is to start with 
the classical Kulback-Leibler information number,
$$D(f\|g)=\int f({\bf x})log (f({\bf x})/g({\bf x}))d{\bf x},$$
where 
$f$ and $g$ are probability densities and ${\bf x}=(x_1, \ldots ,x_n)$.
If $f$ and $g$ are Gaussian with covariance matrices $P$, respectively, $R$,
then
$$D(f\|g)=-\frac{1}{2}\left(\log \det PR^{-1}+\mbox{tr}(I-PR^{-1})\right).$$

By setting, for instance $R = \frac{I}{d}$, we see that $-E (\rho )$ is
upto a constant $D(f\|g)$. Thus, loosely speaking, in this view
the classical analogue of a state is 
a Gaussian whose covariance matrix the density matrix is. 
The foregoing suggests the consideration of
the following entropy of a $d$-dimensional 
state,
$$E(\rho )= \frac{1}{d}\log\det (\rho )=
\frac{1}{d}\sum _{k=1}^d\log \lambda _k,$$
where $\{\lambda _k\}_{k=1}^d$ is the set of eigenvalues of $\rho $.
This is a well-known formula giving a functional of entropy 
type (see \cite{OP}), and also plays a significant role
in convex optimization, \cite{NN}.

$E$ behaves quite differently from the von Neumann entropy, for instance, 
pure states have vanishing von Neumann entropy while $E(\rho )= -\infty $
for a pure state $\rho $ (however, see below for
one variation on $E (\rho )$ which
remedies this). Still, $E$ has some interesting properties, 
two of them described by the following result.

 \begin{theorem}\label{additive}
$(a)$\quad 
Let $\rho $ be a state with parameters $\{g_{k,j}\}$. Then
\begin{equation}\label{prod}
E(\rho  )= \frac{1}{d}\left(\sum _{k=1}^d\log \rho _{k,k}+
\sum _{k<j}\log (1-|g_{kj}|^2)\right).
\end{equation}
$(b)$\quad
Let $\rho $ and $\psi $ be two states. 
Then 
$$E(\rho  \otimes \psi )=E(\rho )+E(\psi ).$$
 \end{theorem}

\begin{proof}
$(a)$  Using Algorithm 10 and Th 11 yields (cf., Th 1.5.10 in \cite{Co})  
$$\det \rho =\left(\prod _{k=1}^d\rho _{k,k}\right)\prod _{k<j}
(1-|g_{kj}|^2).$$
Taking the logarithm in this formula we obtain 
\eqref{prod}.

$(b)$  This follows easily from the formula for the determinant of a tensor
product of matrices. 
\end{proof}

We also notice that $E(\rho )\leq \log d$, with equality for
$\rho _{11}=\ldots =\rho _{dd}=\frac{1}{d}$, which implies the following
maximum entropy principle:
if $\rho $ is a given state, then for any dimension $d'$ there is 
a state $\psi _0$ of dimension $d'$ such that 
$$\max _{\psi }E(\rho \otimes \psi )=E(\psi _0),$$
where the maximum is taken over all states $\psi $ of dimension $d'$.

One small variation on $E$ is provided by the formula:
$$E_0(\rho )=\frac{1}{d}\sum _{j=1}^l\log \lambda _{k_j},$$
where $\lambda _{k_j}$, $j=1,\ldots ,l$, are the nonzero eigenvalues 
of $\rho $.
We see that $E_0$ is still additive and $\rho $ is pure if and only if
$E_0(\rho )=0$. However, to apply \eqref{prod}, one must first find
the restriction of $\rho$ to its support.    

The key utility of using the parameters $g_{kj}$ is that there is no
{\it need} for any eigenvalue computations for finding $E(\rho )$. 
Contrary to the situation with eigenvalues, the $g_{kj}$ can be related
to the entries of $\rho$ via explicit formulae.

\noindent
{\it 3.4. The Peres-Horodecki Criterion.}
The parameters $\{g_{k,j}\}$ of a state $\rho $ can be also
used to 
explicitly write the finite set of inequalities characterizing the 
separability of $2\times 2$ and $2\times 3$ states. For $2\times 2$
states we have the following result.

\begin{theorem}\label{entangle}
Let $\rho $ be a $2\times 2$ state with parameters $\{g_{k,j}\}$.
Then $\rho $ is separable if and only if:
$$
\rho ^{1/2}_{22}\rho ^{1/2}_{33}+
\rho ^{1/2}_{11}\rho ^{1/2}_{44}
d_{12}d_{13}d_{24}d_{34} \geq 
$$
$$ 
\rho ^{1/2}_{11}\rho ^{1/2}_{44}
|g_{12}g_{23}g_{34}+d_{12}g_{13}d_{23}g_{34}+g_{12}d_{23}g_{24}d_{34}
-d_{12}g_{13}\overline{g}_{23}g_{24}d_{34}|
$$
and the system of inequalities
$$d_{12}\left(d_{23}+(1-|h_{14}^2|)^{1/2}\right)
\geq |g_{12}g_{23}-\overline{g}_{12}h_{14}|,$$
$$d_{34}\left(d_{23}+(1-|h_{14}^2|)^{1/2}\right)
\geq |g_{34}g_{23}-\overline{g}_{34}h_{14}|,$$
$$
\rho ^{1/2}_{22}\rho ^{1/2}_{33}+
\rho ^{1/2}_{11}\rho ^{1/2}_{44}
d_{12}(1-|h_{13}^2|)^{1/2}(1-|h_{24}^2|)^{1/2}d_{34} \geq 
$$
$$
\rho ^{1/2}_{11}\rho ^{1/2}_{44}
|\overline{g}_{12}h_{14}\overline{g}_{34}+
d_{12}h_{13}(1-|h_{14}^2|)^{1/2}\overline{g}_{34}+
\overline{g}_{12}(1-|h_{14}^2|)^{1/2}h_{24}d_{34}
-d_{12}h_{13}\overline{h}_{14}h_{24}d_{34}|
$$
admits solutions $h_{14}$, $h_{13}$, $h_{24}$ in the closed unit disk, 
subject to the boundary condition
\eqref{boundary}, where $d_{k,j}=(1-|g_{k,j}|^2)^{1/2}$.
\end{theorem}
\begin{proof}
The result is a consequence of the Peres-Horodecki
characterization of 
$2\times 2$ separable states, \cite{hordecki,Pe}, and Theorem~\ref{param}.
\end{proof}

Similar inequalities can be written for $2\times 3$ separable states.

\section{Quantum channels}
In this section we analyse some consequences of Theorem~\ref {param}
for the structure of quantum channels. We can immediately exemplify the
case of binary channels.

 \begin{example}\label{m2}
 {\rm
 A detailed analysis of quantum binary channels is given in 
 \cite{RSW}. We show here how Theorem~\ref{param} relates to that 
 analysis. It is showed in \cite{KR} that any 
 quantum binary channel $\Phi $ has a representation
 $$\Phi (A)=U[\Phi _{{\bf t}, {\bf \Lambda }}(VAV^*)]U^*,$$
 where $U,V \in U(2)$ and 
 $\Phi _{{\bf t}, {\bf \Lambda }}$
 has the matrix representation
 $${\bf T}=\left[\begin{array}{cccc}
 1 & 0 & 0 & 0 \\
 t_1 &  \lambda _1 & 0 & 0 \\
 t_2 & 0 & \lambda _2 & 0 \\
 t_3 & 0 & 0 & \lambda _3 
 \end{array}\right]$$
 with respect to the Pauli basis 
 $\{I,\sigma _1,\sigma _2,\sigma _3\}$ of $\cM_2$.
See \cite{vvi} also for similar normal forms.
We can obtain (formula (26) in \cite{RSW}) that
 $$S_{\Phi _{{\bf t}, {\bf \Lambda }}}=
 \frac{1}{2}\left[\begin{array}{cccc}
 1+t_3+\lambda _3 & t_1-it_2 & 0 & \lambda _1+\lambda _2 \\
 t_1+it_2 & 1-t_3-\lambda _3 & \lambda _1-\lambda _2 & 0 \\
 0 & \lambda _1-\lambda _2 & 1+t_3-\lambda _3 & t_1-it_2 \\
 \lambda _1+\lambda _2 & 0 & t_1+it_2 & 1-t_3+\lambda _3
 \end{array}\right].
 $$
 Similarly, by formula (27) in \cite{RSW}, 
 $$S_{\hat{\Phi }_{{\bf t}, {\bf \Lambda }}}=
 \frac{1}{2}\left[\begin{array}{cccc}
 1+t_3+\lambda _3 & 0 & t_1+it_2 & \lambda _1+\lambda _2 \\
 0 & 1+t_3-\lambda _3 & \lambda _1-\lambda _2 & t_1+it_2 \\
 t_1-it_2 & \lambda _1-\lambda _2 & 1-t_3-\lambda _3 & 0 \\
 \lambda _1+\lambda _2 & t_1+it_2 & 0 & 1-t_3+\lambda _3
 \end{array}\right].
 $$
 It is slightly more convenient to deal with 
 $S=[S_{k,j}]_{k,j=1}^4=2S_{\hat{\Phi }_{{\bf t}, {\bf \Lambda }}}$.
 Formula \eqref{relation} gives:
 $$S_{11}= 1+t_3+\lambda _3;\quad \quad 
 S_{22}= 1+t_3-\lambda _3;$$
 $$S_{33}= 1-t_3-\lambda _3;\quad \quad 
 S_{44}= 1-t_3+\lambda _3;$$
 $$\Gamma _{12}=0, \quad \Gamma _{34}=0;$$
 $$S_{23}= S _{22}^{1/2}\Gamma _{23}S_{33}^{1/2},$$
 so that 
 $$\Gamma _{23}=\frac{\lambda _1-\lambda _2}{(1+t_3-\lambda _3)^{1/2}
 (1-t_3-\lambda _3)^{1/2}};$$
 $$S_{13}=S^{1/2}_{11}\Gamma _{13}D_{\Gamma _{23}}
 S^{1/2}_{33},$$
 so that 
 $$\Gamma _{13}=\frac{(t_1+it_2)(1+t_3-\lambda _3)^{1/2}}
 {((1+t_3-\lambda _3)
 (1-t_3-\lambda _3)-(\lambda _1-\lambda _2)^2)^{1/2}
 (1+t_3+\lambda _3)^{1/2}};$$
 $$S_{24}= S^{1/2}_{22}D_{\Gamma ^*_{23}}\Gamma _{24}
 S^{1/2}_{44},$$
 so that 
 $$\Gamma _{24}=\frac{(t_1+it_2)(1-t_3-\lambda _3)^{1/2}}
 {((1+t_3-\lambda _3)
 (1-t_3-\lambda _3)-(\lambda _1-\lambda _2)^2)^{1/2}
 (1-t_3+\lambda _3)^{1/2}}.$$
 Finally, 
 $$S_{14}= S^{1/2}_{11}(-\Gamma _{13}\Gamma ^*_{23}\Gamma _{24}+
 D_{\Gamma ^*_{13}}\Gamma _{14}D_{\Gamma _{24}})S^{1/2}_{44}.$$
 For brevity, we omit writing out the formula for $\Gamma _{14}$. 
 We deduce that  ${\Phi }_{{\bf t}, {\bf \Lambda }}$ is completely positive 
 if and only if the following eight inequalities hold:
 $$S _{k,k}\geq 0, \quad k=1,\ldots ,4,$$
 $$|\Gamma _{23}|\leq 1,\quad |\Gamma _{13}|\leq 1,\quad
 |\Gamma _{24}|\leq 1,\quad |\Gamma _{14}|\leq 1.$$
 Further, we know what happens in the degenerate cases (i.e., the cases where
any of these inequalities become equalities).
Thus, the implication of $S_{kk}=0$ for some $k$ 
on the structure of ${\Phi }_{{\bf t}, {\bf \Lambda }}$ is clear.
 Also, if $|\Gamma _{23}|=1$, 
 then necessarily $t_1=t_2=0$ and 
 $\lambda _1+\lambda _2=(1+t_3+\lambda _3)^{1/2}\Gamma _{14}
 (1-t_3+\lambda _3)^{1/2}$ for some contraction $\Gamma _{14}$.
 If either $|\Gamma _{13}|=1$ or $|\Gamma _{24}|=1$, then 
 necessarily 
$\Gamma _{14}=0$ and 
 $S_{14}= S^{1/2}_{11}
(-\Gamma _{13}\Gamma ^*_{23}\Gamma _{24})S^{1/2}_{44}$.

 We notice that this result is of about the same nature as that 
 in \cite{RSW}. This is because the first step 
 of \eqref{relation}, specialized to $4\times 4$ matrices
( viz., the $j = k + 1$ step),
 is precisely Lemma~6 in \cite{RSW} which is 
 used for the analysis in \cite{RSW}. If we had instead used the block version 
 of \eqref{relation}, i.e., viewing $S_{\Phi}$ as a $2\times 2$ matrix with
entries themselves $2\times 2$ matrices,
then we would deduce precisely Theorem~1
 of \cite{RSW}. What we basically have done here is that we used
 \eqref{relation} in order to deduce in a systematic way 
 the condition that $R_{{\Phi }_{{\bf t}, {\bf \Lambda }}}$
 in Theorem~1 of \cite{RSW} is a contraction. One advantage of 
 doing this is that it works in higher dimensions.

It is noted that the correspondence between $S_{\Phi }$ and 
the parameters $\Gamma _{k,j}$ is nonlinear. Only for the first
step is the correspondence affine and therefore can be used in the 
analysis of extreme points in the case $d=2$, as it was done in \cite{RSW}. 
This seems to be 
unclear for $d > 2$, at this moment. 
 }\qed
 \end{example}

\noindent
{\it 4.1. Capacity.}
It is worth remarking that the suggested notion of quantum entropy
from Section (3.3) could be used to posit a notion of channel capacity
for quantum channels, which is trivially additive.
Several numbers have been suggested to define the capacity of a quantum 
channel, with additivity as a desirable requirement. See, for instance,
\cite{shor,WH}.
While additivity of some these notions has been demonstrated for special
classes of channels, such as entanglement breaking channels, conjectures in the 
direction of additivity for some other notions
were recently disproved (\cite{WH}).
Therefore, it
might be useful to have some other possible candidates.
Motivated by the discussion in Section ~3.3 we introduce for a quantum 
channel
$\Phi :{\cM }_d\rightarrow {\cM }_{d'}$ the number
$$D(\Phi )=-\frac{1}{N}\log \det S_{\Phi },
$$
where $N=dd'$. Trivially, $D(\Phi )$ is additive, and there is an explicit
expression for it in terms of the parameters $\Gamma_{kj}$ of $S_{\Phi}$: 
\begin{equation}\label{bprod}
D(\Phi )=-\frac{1}{N}\left(\sum _{k=1}^N\log S_{k,k}+
\sum _{k<j}\log (1-|\Gamma _{k,j}|^2)\right).
\end{equation}
$(b)$\quad

Once again, $D(\Phi )\geq \log n$ and a minimum capacity principle holds:
given $\Phi $, there exists a quantum channel 
$\Psi $ such that 
$$\min _{\Psi }D(\Phi \otimes \Psi )$$
is attained.
As an example of explicit computation of $D(\Phi )$, we consider 
binary quantum channels as in Example \ref{m2}. 
For simplicity, assume $t_1=t_2=t_3=0$. Then, using the 
parameters calculated in Example \ref{m2},  
we deduce
$$\begin{array}{rcl}
D(\Phi ) &=& -1/4\left(
2\log \left(\frac{1+\lambda _3}{2}\right)+ 
2\log \left(\frac{1-\lambda _3}{2}\right) \right. \\ 
 & & \\
& & \quad \left.
+\log \left(1-|\frac{\lambda _1-\lambda _2}{1-\lambda _3}|^2\right)+
\log \left(1-|\frac{\lambda _1+\lambda _2}{1+\lambda _3}|^2\right)
\right).
\end{array}
$$

\noindent
{\it 4.2. Connections with Kraus representations.}
In this subsection, we provide one explicit Kraus operator
representation for any channel, which {\it does not} require
either eigenvalues or eigenvectors of the associated Choi matrix.
While Kraus representations are not unique, they are very useful in
explicit computations involving quantum channels. For instance, one can 
write down at least one Stinespring
dilation in terms of a Kraus representation, and thus at least one
mock unitary operation for a quantum channel; a formula for entanglement
fidelity of a channel can be computed in terms of them; sufficient conditions
for either unitarity or the entanglement breaking property can be checked;
they play a role in quantum error correcting codes, quantum tomography etc.,
See, for instance, \cite{nie,preskill,tim}. 
\vspace*{0.9mm}

\noindent Consider a quantum channel $\Phi :\cM _d\rightarrow \cM _d$.
A familiar representation of quantum channels is the Kraus 
representation, \cite{Kr},
$$\Phi (\rho )=\sum _{k=1}^{d^2}A^*_k\rho A_k,$$
where $A_k\in \cM _d$, $k=1,\ldots ,d^2,$ are called the generators of
$\Phi $, and $\sum _{k=1}^{d^2}A_kA^*_k=I$. The connection 
between the Kraus representation and $S_{\Phi }$ is given by the formula
\begin{equation}\label{con}
S_{\Phi }=A^*A,
\end{equation}
where 
$$A=\left[\begin{array}{c}
\mbox{row}(A_1) \\
\vdots \\
\mbox{row}(A_{d^2})
\end{array}
\right]
$$
and 
$$\mbox{row}(A_k)=\left[\begin{array}{cccccccccc}
(A_k)_{11} & \ldots & (A_k)_{1d} & (A_k)_{2d} & \ldots & (A_k)_{2d} &
\ldots & (A_k)_{d1} & \ldots & (A_k)_{dd} 
\end{array}
\right].
$$

\noindent
This relation shows that while the generators of a Kraus
representation are not unique, they will give the same 
channel provided that $\eqref{con}$ holds. 
{\it In particular, we can choose Kraus generators using the Cholesky
factorization of $S_{\Phi }$}. As mentioned before, the algorithms for
finding the parameters, $\Gamma_{kj}$, systematically yield a Cholesky
factorization of a positive matrix.  
We give the details for binary channels.
\begin{example}\label{kraus}
 {\rm Let $S_{\Phi }:\cM _2\rightarrow \cM _2$ be a binary channel
and let $\{\Gamma _{k,j}\}$ be the parameters of $\Phi $ given by
Theorem~ \ref{param}. Using formula \eqref{cholescu}, we deduce
$$G_{14}=\left[
\begin{array}{cccc}
L_{11} & \Gamma _{12}L_{22} & ZL_{33} & WL_{44} \\
0  & D_{\Gamma _{12}}L_{22} & 
\left(D_{\Gamma _{12}}\Gamma _{23}-\Gamma ^*_{12}\Gamma _{13}D_{\Gamma _{23}}
\right)L_{33} & XL_{44} \\
0 & 0 & D_{\Gamma _{13}}D_{\Gamma _{23}}L_{33} & YL_{44} \\
0 & 0 & 0 & D_{\Gamma _{14}}D_{\Gamma _{24}}D_{\Gamma _{34}}L_{44}
\end{array}\right],
$$
with 
$$\begin{array}{rcl}
X&=&D_{\Gamma _{12}}\Gamma _{23}\Gamma _{34}+  
D_{\Gamma _{12}}D^*_{\Gamma _{23}}\Gamma _{24}D_{\Gamma _{34}} \\
 & & \\
 & & \quad -\Gamma ^*_{12}\Gamma _{13}D_{\Gamma _{23}}\Gamma _{34}
+\Gamma ^*_{12}\Gamma _{13}\Gamma ^*_{23}\Gamma _{24}D_{\Gamma _{34}}
-\Gamma ^*_{12}D_{\Gamma ^*_{13}}\Gamma _{14}D_{\Gamma _{24}}D_{\Gamma _{34}},
\end{array}
$$
$$Y=D_{\Gamma _{13}}D_{\Gamma _{23}}\Gamma _{34}-
D_{\Gamma _{13}}\Gamma ^*_{23}\Gamma _{24}D_{\Gamma _{34}}
-\Gamma ^*_{13}\Gamma _{14}D_{\Gamma _{24}}D_{\Gamma _{34}},
$$
$$Z=\Gamma _{12}\Gamma _{23}+
D_{\Gamma ^*_{12}}\Gamma _{13}D_{\Gamma _{23}},
$$
and 
$$\begin{array}{rcl}
W&=&\Gamma _{12}\Gamma _{23}\Gamma _{34}+
D_{\Gamma ^*_{12}}\Gamma _{13}D_{\Gamma _{23}}\Gamma _{34}
+\Gamma _{12}D_{\Gamma ^*_{23}}\Gamma _{24}D_{\Gamma _{34}} \\
 & & \\
 & & \quad -D_{\Gamma ^*_{12}}\Gamma _{13}\Gamma ^*_{23}\Gamma _{24}
D_{\Gamma _{34}}+D_{\Gamma ^*_{12}}D_{\Gamma ^*_{13}}
\Gamma _{14}D_{\Gamma _{24}}D_{\Gamma _{34}}.
\end{array}
$$
The Kraus generators are
$$A_1=\left[\begin{array}{cc}
L_{11} & \Gamma _{12}L_{22} \\
ZL_{33} & WL_{44}
\end{array}\right],\quad 
A_2=\left[\begin{array}{cc}
0  & D_{\Gamma _{12}}L_{22} \\ 
\left(D_{\Gamma _{12}}\Gamma _{23}-\Gamma ^*_{12}\Gamma _{13}D_{\Gamma _{23}}
\right)L_{33} & XL_{44} 
\end{array}\right],
$$
$$A_3=\left[\begin{array}{cc}
0 & 0 \\
 D_{\Gamma _{13}}D_{\Gamma _{23}}L_{33} & YL_{44} 
\end{array}\right],\quad 
A_4=\left[\begin{array}{cc}
0 & 0 \\ 
0 & D_{\Gamma _{14}}D_{\Gamma _{24}}D_{\Gamma _{34}}L_{44}
\end{array}\right].
$$
}
\end{example}

The advantage is that, once the 
parameters of $S_{\Phi }$ associated by Theorem ~\ref{param} are 
known, the Cholesky factorization can be computed by \eqref{cholescu}.
Indeed, the algorithms in the appendix for calculating the $\Gamma_{kj}$,
systematically compute a Cholesky factorization.
Also, the orthogonality properties of the Cholesky factorization might be 
useful in some situations.
These constructions make sense in infinite dimensions, except that further
qualifications are needed for the trace-preserving property to 
be meaningful.

\section{Appendix}

\noindent In this appendix, a sketch of the proof of Theorem ~\ref{param}
is first given. This proof does not immediately yield a constructive 
procedure for finding the $\Gamma_{kj}$. For that purpose, we next
discuss briefly the notion of displacement structure (which works
for a family of matrices with operator entries). A key feature of a  
displacement structure is that all possible Schur complements of such
a family of matrices inherit a related displacement structure (this is
what enables the determination of Cholesky factorizations of each member
of the family). This is stated in Algorithm \ref{algorithm}.
We next show how any positive matrix can be imbedded into a family of
matrices with displacement structure. Typically, a family with displacement
structure admits more than one such representation. We choose one
such representation - the specific choice for $F(-t)$ enables the
association of a lattice structure to the corresponding version of
Algorithm \ref{algorithm}. Now, if $\cH $ is of finite dimension,
this specific choice of displacement structure made for positive matrices,
results in a significant simplification of Algorithm \ref{algorithm}.
This is stated in Algorithm~\ref{blgorithm}.
In particular, a series of explicitly determined finite-dimensional
contractions, 
$\gamma_{k}(t)$, is produced by the algorithm.
In terms of these, $\gamma_{k}(t)$,
there is a simple and explicit formula for
the parameters, $\Gamma_{kj}$. Thus, the determination of the parameters,
$\Gamma_{kj}$, is fully constructive when $\cH $ is finite dimensional 
(much of the procedure survives even when $\cH $ is
infinite dimensional). 

\vspace*{0.9mm}

\noindent{\it 5.1. Parametrization of positive matrices.}
We now sketch a proof of Theorem ~\ref{param}. Since the diagonal entries
of $S$ intervene in the parametrization only through the
$L_{kk}$ ($S_{kk} = L_{kk}{*}L_{kk}$), we assume, 
without loss of generality, that $S_{k,k}=I_{\cH }$ for all 
$k=1,\ldots ,d$. By a result of Kolmogorov (Theorem ~1.3.1 in \cite{Co}),
there exist a Hilbert space $\cK $ containing $\cH$and isometries
$V_k\in \cL (\cH,\cK )$, $k=1,\ldots ,d$, such that 
$$(i)\,\,\,\mbox{The set $\{V_k\cH \mid k=1,\ldots d\}$ is total in $\cK $};
$$
$$(ii)\,\,\, S_{k,j}=V^*_kV_j, \quad k,j=1,\ldots ,d.$$
Since $V_1$ is an isometry, we can identify $\cH $ with the range of $V_1$ 
and assume $V_1=P^{\cK }_{\cH }/\cH $, where $P^{\cK }_{\cH }$
denotes the orthogonal projection of $\cK $ onto $\cH $.
Since $V_2$ is an isometry, there is a Hilbert space $\cD _2$ and an 
operator $D_2\in \cL (\cD _2,\cK )$ such that 
$$W_1=\left[\begin{array}{cc}
V_2 & D_2
\end{array}\right]:\cH \oplus \cD _2\rightarrow \cK 
$$
is a unitary operator and $V_2=W_1/\cH $. Set $\cK _1=\cK $,
$\cK _2=\cH\oplus \cD _2$. By induction we obtain a family of unitary
operators $W_1\in \cL (\cK _2,\cK _1)$, $\ldots $, 
$W_{d-1}\in \cL (\cK _d,\cK _{d-1})$ such that 
$$V_k=W_1\ldots W_{k-1}/\cH , \quad k=2, \ldots ,d.$$
In particular, 
\begin{equation}\label{coef}
S_{k,j}=P^{\cK }_{\cH }W_k\ldots W_{j-1}/\cH , \quad k<j.
\end{equation}

Now, we notice that by a suitable identification of spaces, 
$$W_1=\left[\begin{array}{ccc}
\Gamma _{12} & a & b \\
D_{\Gamma _{12}} & A_{11} & A_{12} \\
0 & A_{21} & A_{22} 
\end{array}\right].$$
Multiplying this operator on the left by the unitary operator
$R(\Gamma ^*_{12})\oplus I$, we obtain a unitary operator with matrix
representation 
$$\left[\begin{array}{ccc}
I & X & Y \\
0 & A'_{11} & A'_{12} \\
0 & A'_{21} & A'_{22} 
\end{array}\right].$$
It follows that necessarily, $X=0$ and $Y=0$. Iterating this construction 
and using \eqref{coef} and $(ii)$, we deduce 
\eqref{relation}.

\noindent
{\it 5.2. Algorithms} 
As discussed above, the notion of displacement structure is useful in
producing an algorithmic procedure for finding the $\Gamma_{kj}$.
The systematic study of the displacement structure was
initiated in \cite{KKM}. A main theme of the theory 
is the recursive factorization of matrices with implicit
structure encoded by a so-called displacement equation
of the form 
\begin{equation}\label{qaz}
R(t)-F(t)R(t-1)F(t)^*=G(t)J(t)G(t)^*, \quad t\in \ZZ .
\end{equation}
Given $R(t-1)$, we need to know $\{F(t),G(t),J(t)\}$ in order to 
determine $R(t)$ using \eqref{qaz}. We shall not 
seek to determine $R(t)$ by explicitly applying \eqref{qaz}. 
Instead, we shall use the fact that $R(t)$ is a ``low rank'' 
modification of $R(t-1)$ and exploit it to determine $R(t)$ 
more efficiently. This idea will operate as follows. Use as 
input data $\{F(t),G(t),J(t)\}$ and the Cholesky factor 
of $R(t-1)$, say $\bar{L}(t-1)$, and then compute the 
Cholesky factor of $R(t)$ without  determining 
$R(t)$,
\begin{equation}
R(t)=\bar{L}(t)\bar{L}^*(t)\;.
\end{equation}
The following algorithm, see \cite{CSK}, tells us how 
to compute the columns of $\bar{L}(t)$ from the 
columns of $\bar{L}(t-1)$ and knowledge of $\{F(t),G(t),J(t)\}$. 
Here we describe the algorithm for special types of $\{F(t),G(t),J(t)\}$,
which subsume the case of positive-definite 
matrices $R(t)$.

Let $\bar{l}_k(t)$ denote the nonzero part of the $k-$th column of 
$\bar{L}(t)$. Let also $1/\sqrt{d_k(t)}$ denote the top entry of 
$\bar{l}_k(t)$ and define
$$
l_k(t)=\sqrt{d_k(t)}\;\bar{l}_k(t).
$$
That is, the top entry of $l_k(t)$ is normalized to $1$, and we also 
obtain the equivalent triangular factorization for $R(t)$,
\begin{equation}\label{triang.fact} 
R(t)=L(t)D^{-1}(t)L^*(t)\;,
\end{equation}
where the diagonal entries of $D(t)$ are the $\{d_k(t)\}$ and 
the columns of $L(t)$ are the $\{l_k(t)\}$. Here, $L(t)$ is a
unit diagonal lower triangular matrix.
Let $R_k(t)$ denote the Schur complement of $R(t)$ with 
respect to its leading $k\times k$ block.
{\it Suppose} $F(t)$ is lower triangular with diagonal entries $f_k(t)$
and let $F_k(t)$ denote the matrix obtained by deleting 
the first $k$ rows and columns of $F(t)$. 

\begin{algorithm} \label{algorithm}
Assume we know 
$$\{L(t-1),D(t-1),G(t),F(t),J(t)\}\;,
$$
and that $R(t)$ satisfies the 
displacement equation (\ref{qaz}). 
Then, for each $k$, the Schur complements, $R_{k}(t)$,
of $R(t)$ with respect to its leading $k\times k$ block, satisfy similar 
displacement equations,
\begin{equation}\label{as.disp.sch}
 R_k(t)-F_k(t)R_k(t-1)F_k^*(t)=G_k(t)J(t)G_k^*(t),
\end{equation}
where the $G_k(t)$, and the triangular factorization (\ref{triang.fact}) 
of 
$R(t)$, can be obtained from the following 
recursive construction:

\begin{equation}\label{SA}
\left[\begin{array}{cc}
   & 0 \\ l_k(t) & \\    & G_{k+1}(t)
\end{array}\right]=
\left[\begin{array}{cc}
F_k(t)l_k(t-1) & G_k(t)
\end{array}\right]
\left[\begin{array}{cc}
 f^*_k(t) & h_k^*(t)\Theta_k(t) \\
 & \\
J(t)g^*_k(t) & p^*_k(t)\Theta_k(t) 
\end{array}\right],
\end{equation}
where $g_k(t)$ is the top  row of $G_k(t),$  and $h_k(t)$ and 
 $p_k(t)$ are chosen so as to satisfy the relation
$$
\left[\begin{array}{cc}
f_k(t) & g_k(t) \\
h_k(t) & p_k(t)
\end{array}
\right]
\left[\begin{array}{cc}
d_k(t-1) & 0 \\
 0 & J(t)
\end{array}
\right]
\left[\begin{array}{cc}
f_k(t) & g_k(t) \\
h_k(t) & p_k(t)
\end{array}
\right]^*
=
\left[\begin{array}{cc}
d_k(t) & 0 \\
 0 & J(t)
\end{array}
\right].
$$

\end{algorithm}

It is shown in \cite{CSK} that it is always possible to 
find $h_k(t)$ and $p_k(t)$ as above. Choices  
that result in array (lattice) form descriptions  
are also possible and are described in detail in \cite{CSK}.

\noindent {\it Displacement Structure for Positive Matrices}
Let $S$ be positive. 
For $t=0,\ldots ,d-1$,
$$R(-t)=\left[
\begin{array}{ccccc}
I & S_{t+1,t+2} & & & S_{t+1,d} \\
S^*_{t+1,t+2} & I & & & S_{t+2,d} \\ 
 & & \ddots & & \\ 
 & & & I & S_{d-1,d} \\
S^*_{t+1,d} & & & S_{d-1,d}^* & I
\end{array}\right]\oplus \underbrace{I\oplus \ldots \oplus I}_{t \,\,times},
$$
so that, $R(0)=S_{\Phi }$ and 
$R(-d+1)=\underbrace{I\oplus \ldots \oplus I}_{d \,\,times}$.
Then, for $t=0, \ldots, d-1$,
$$
F(-t)=F=\left[\begin{array}{ccccc}
0 & & & & \\
I & 0 & & & \\
 & & \ddots & & \\
 & & & 0 & \\
 & & & I & 0
\end{array}\right], \quad 
G(-t)=\left[\begin{array}{cc}
I & 0 \\
S^*_{t+1,t+2} & S^*_{t+1,t+2} \\
 \vdots & \vdots \\
S^*_{t+1,d} & S^*_{t+1,d} \\ 
0 & 0 \\
\vdots & \vdots \\
0 & 0 
\end{array}\right],
$$
and
$$
J(-t)=J=\left[\begin{array}{cc}
I & 0 \\
0 & -I 
\end{array}
\right].
$$
We easily check that 
\begin{equation}\label{poz}
R(-t)-FR(-t-1)F^*=G(-t)JG(-t)^*, 
\quad t=1,\ldots ,d-2,
\end{equation}
so that, $d-2$ steps of Algorithm \ref{algorithm}
will exploit the whole information encoded in 
$R(0)=S_{\Phi }$.

We notice that Algorithm \ref{algorithm}
is greatly simplified with the special choice of $R(t), F(t), G(t)$,
when $\cH $ is of finite dimension. 
Thus, write $g_k(t)=\left[u_k(t) \,\,\,v_k(t)\right]$
with respect to the decomposition 
of $J$ and since $f_k(t)=0$, we deduce
$d_k(t)=g_k(t)Jg_k(t)^*=
u_k(t)u_k(t)^*-v_k(t)v_k(t)^*$.
The positivity of $S_{\Phi }$ is therefore tested by 
the condition $u_k(t)u_k(t)^*-v_k(t)v_k(t)^*\geq 0$
for all $t$ and all $k$. If this holds for some $t$ and some $k$, 
then there exist {\it explicitly} determined contractions,  
$\gamma _k(t)$, satisfying such that $v_k(t)=u_k(t)\gamma _k(t)$. 
In this situation we introduce
$$\Theta _k(t)=\left[\begin{array}{cc}
I & -\gamma _k(t) \\
-\gamma _k(t)^* & I
\end{array}
\right]
\left[\begin{array}{cc}
(I-\gamma _k(t)\gamma _k(t)^*)^{-1/2} & 0 \\
0 & (I-\gamma _k(t)^*\gamma _k(t))^{-1/2}
\end{array}
\right]
$$
and then the generator recursion in Algorithm~\ref{algorithm}
reduces to a simpler form.

\begin{algorithm}\label{blgorithm}
The generator recursion for the displacement equation
\eqref{poz} associated to a positive matrix $S $
has the form:
\begin{equation}\label{gener}
\left[
\begin{array}{c}
0 \\
G_{k+1}(t)
\end{array}
\right]=
FG_k(t-1)\Theta _k(t-1)\left[
\begin{array}{cc}
I & 0 \\
 0 & 0
\end{array}
\right]+
G_k(t)\Theta _k(t)\left[
\begin{array}{cc}
 0 & 0 \\
 0 & I
\end{array}
\right].
\end{equation}
\end{algorithm}

Algorithm \ref{blgorithm}
produces an explicit connection between 
the parameters $\gamma _k(t)$ and the entries
$S_{k,j}$ of $S$. Namely, a direct calculation
yields the following result.

\begin{theorem}\label{final}
$\Gamma _{k,j}=\gamma _{j-k}(1-k)^*$, $1\leq k<j\leq d$.
\end{theorem}

\end{document}